\documentclass[a4paper]{siamart1116}

\usepackage{dsfont}
\usepackage{amsmath}
\usepackage{amssymb}

\usepackage{amsfonts}
\usepackage{graphicx}
\usepackage{epstopdf}
\usepackage{algorithmic}
\Crefname{ALC@unique}{Line}{Lines} % <- Preamble
\usepackage{caption}
\usepackage{upquote}

\ifpdf
  \DeclareGraphicsExtensions{.eps,.pdf,.png,.jpg}
\else
  \DeclareGraphicsExtensions{.eps}
\fi

\usepackage{geometry}
\numberwithin{theorem}{section}

\newcommand{\TheTitle}{On Checking Null Rank Conditions of Rational Matrices}
\newcommand{\TheAuthors}{A. Varga}

\headers{\TheTitle}{\TheAuthors}

\title{{\TheTitle}}%\thanks{Submitted to the editors DATE.
\author{
  Andreas Varga\thanks{Gilching, Germany
    (\email{varga.andreas@gmail.com}).}
}
\headers{On Checking Null Rank Conditions of Rational Matrices}{A. Varga}

\ifpdf
\hypersetup{
  pdftitle={\TheTitle},
  pdfauthor={\TheAuthors}
}
\fi

\newcommand{\be}{\begin{equation}}
\newcommand{\ee}{\end{equation}}
\newcommand{\ba}{\left [ \begin{array}}
\newcommand{\ea}{\end{array} \right ]}
\newcommand{\bea}{\begin{eqnarray}}
\newcommand{\eea}{\end{eqnarray}}

\newcommand{\rank}{\mathop{\mathrm{rank}}}

\newcommand{\image}{\mathop{\mathrm{Im}}}

\begin{document}

\maketitle

% REQUIRED
\begin{abstract}
In this paper we discuss possible numerical approaches to reliably check the rank condition $\rank G(\lambda) = 0$ for a given rational matrix $G(\lambda)$ in terms of its descriptor system realization. For test purposes we employ functions available in the Control System Toolbox of MATLAB and the Descriptor System Tools (DSTOOLS) collection.
\end{abstract}

% REQUIRED
\begin{keywords}
  rational matrices, computational methods, descriptor systems
\end{keywords}

% REQUIRED
\begin{AMS}
   	26C15, 93B40, 93C05, 93B55
\end{AMS}

\section{Introduction}

In this section we present some basic notions on descriptor systems, and describe the main use of testing null rank conditions of rational matrices.

Let $G(\lambda)$ be a $p\times m$ real rational matrix  and consider an $n$-th order descriptor system realization $(A-\lambda E,B,C,D)$, with $A-\lambda E$ an $n\times n$ regular pencil (i.e., $\det(A-\lambda E) \not\equiv 0$), which satisfies
\be\label{GTFM} G(\lambda) = C(\lambda E-A)^{-1}B+D .\ee
We will also use the equivalent notation for the TFM (\ref{GTFM})
\be\label{GTFMalt} G(\lambda) = \ba{c|c} A-\lambda E & B\\ \hline C & D \ea .\ee
If $Q, Z \in \mathds{R}^{n\times n}$ are invertible matrices, then the two realizations $(A-\lambda E,B,C,D)$ and $(\widetilde A-\lambda \widetilde E,\widetilde B,\widetilde C,D)$, whose matrices are related by a similarity transformation of the form
\be\label{dssim} \widetilde A-\lambda \widetilde E = Q(A-\lambda E)Z, \quad \widetilde B = QB, \quad \widetilde C = CZ \, ,\ee
have the same TFM $G(\lambda)$.

The rational matrix $G(\lambda)$ in (\ref{GTFM}) can be seen as the transfer function matrix of a generalized linear time-invariant system with $m$ inputs and $p$ outputs.
According to the system type, the frequency variable $\lambda$ is either $\lambda = s$, the complex variable in the Laplace-transform
  in the case of a continuous-time system or $\lambda = z$,
the complex variable in the Z-transform in the case of a
discrete-time system.

We recall from \cite{Verg79,Verg81} some basic notions related to descriptor system realizations.  A  realization $(A-\lambda E,B,C,D)$ of a rational matrix $G(\lambda)$ is called \emph{minimal} if it is controllable, observable and has no non-dynamic modes. For a minimal realization, $n$ has its least possible value and all minimal realizations have the same order $n$. The poles of $G(\lambda)$ are related to $\Lambda(A-\lambda E)$, the eigenvalues of the \emph{pole pencil}  $A-\lambda E$ (also known as the generalized eigenvalues of the pair $(A,E)$).
For a minimal  realization, the finite poles of $G(\lambda)$ are the finite eigenvalues of $A-\lambda E$, while the multiplicities of infinite poles are defined by the multiplicities of infinite eigenvalues of $A-\lambda E$ minus one. The infinite eigenvalues of multiplicity ones are the so-called non-dynamic modes. The McMillan degree of $G(\lambda)$, denoted by $\delta\big(G(\lambda)\big)$, is the number of poles of $G(\lambda)$, finite and infinite, counting all multiplicities.

The minimality property involves the controllability and observability properties of the descriptor realization $(A-\lambda E,B,C,D)$.
A finite eigenvalue $\lambda_f \in \Lambda(A-\lambda E)$ is controllable if $\rank\,[\,A-\lambda_f E \; B\,] = n$, otherwise is uncontrollable. The pair $(A-\lambda E,B)$ is finite controllable if all finite eigenvalues in $\Lambda(A-\lambda E)$ are controllable.  Similarly, a finite eigenvalue $\lambda_f \in \Lambda(A-\lambda E)$ is observable if $\rank\,[\,A^T-\lambda_f E^T \; C^T\,] = n$, otherwise is unobservable. The pair $(A-\lambda E,C)$ is finite observable if all finite eigenvalues in $\Lambda(A-\lambda E)$ are observable. Infinite controllability requires that $\rank\,[\, E \; B\,] = n$, while infinite observability requires that $\rank\,[\,E^T \; C^T\,] = n$. The realization $(A-\lambda E,B,C,D)$ is controllable if the pair $(A-\lambda E,B)$ is  both finite and infinite controllable. Similarly, the realization $(A-\lambda E,B,C,D)$ is observable if the pair $(A-\lambda E,C)$ is both finite and infinite observable. The third requirement for minimality, the lack of non-dynamic modes, can be equivalently expressed as $A \ker(E) \subseteq \image(E)$.  This characterization of minimality has been given in \cite{Verg79}.

Assume the descriptor system (\ref{GTFMalt}) is minimal. We say the descriptor system (\ref{GTFMalt}) is \emph{proper} if  all infinite eigenvalues of the pole pencil $A-\lambda E$ are simple. For a descriptor realization (\ref{GTFMalt}) which is potentially non-minimal, the definition of properness can be extended as follows. We say the descriptor system (\ref{GTFMalt}) is \emph{proper} if the infinite eigenvalues of the pole pencil of its infinite controllable and infinite observable realization  are simple.

The rank, or more precisely the \emph{normal rank}, of a rational matrix $G(\lambda)$, which we also denote by $\rank G(\lambda)$, is the maximal number of linearly independent rows (or columns) over the field of rational functions $\mathds{R}(\lambda)$.
It can be shown that the normal rank of $G(\lambda)$ is the maximally possible rank of the complex matrix $G(\lambda)$ for all values of $\lambda \in \mathds{C}$ such that $G(\lambda)$ has finite norm. This interpretation provides a simple way to determine the normal rank as the maximum of the rank of $G(\lambda)$ for a few random values of the frequency variable $\lambda$.

Checking the rank condition $\rank G(\lambda) = 0$ or equivalently $G(\lambda) = 0_{p\times m}$ (i.e., the $p\times m$ null matrix) appears frequently when verifying the results of various computations involving rational matrices. For example, the computed inverse $\overline R(\lambda) = R^{-1}(\lambda)$  of an invertible  rational matrix $R(\lambda)$ must trivially satisfy $G(\lambda) = 0$, where $G(\lambda) = \overline R(\lambda) R(\lambda)-I$. Also, for an arbitrary rational matrix $R(\lambda)$ the resulting factors $N(\lambda)$ and $M(\lambda)$ of any rational factorization representation $R(\lambda) = M^{-1}(\lambda)N(\lambda)$, must satisfy $G(\lambda) = 0$, for $G(\lambda) = M(\lambda) R(\lambda)-N(\lambda)$. And finally, the rational left  nullspace basis $N_l(\lambda)$ of a rational matrix $R(\lambda)$ must satisfy $G(\lambda) = 0$, for $G(\lambda) = N_l(\lambda)R(\lambda)$.  See \cite[Ch. 9]{Varg17} for many other potential examples. A it is apparent from these examples, is that typically the descriptor realization of $G(\lambda)$ is non-minimal and may contain uncontrollable and unobservable eigenvalues (both finite and infinite) as well as non-dinamic modes.

In what follows, we review different approaches for checking of rank condition $\rank G(\lambda) = 0$ in the light of available software in the Control System Toolbox of MATLAB \cite{MLCO18} and in the Descriptor System Tools (\textbf{DSTOOLS}) developed by the author \cite{Varg18}. Simple examples are used to illustrate the capabilities and limitations of the numerical methods underlying the existing software implementations.

\section{Review of methods to check $\rank G(\lambda) = 0$}\label{sec:methods}
In the following, we discuss several methods which are suited to check the rank condition $\rank G(\lambda) = 0$, using the a descriptor realization (\ref{GTFMalt}) of $G(\lambda)$, for which no minimality is assumed. For each approach we shortly discuss the underlying computational, discuss the advantages and limitations of each method and indicate available software implementations.

\subsection{Method 1: Computing the normal rank of a minimal realization}
A straightforward approach to check the null rank condition $\rank G(\lambda) = 0$ is to compute a minimal realization $(A_m-\lambda E_m,B_m,C_m,D_m)$ starting from a non-minimal realization $(A-\lambda E,B,C,D)$ of $G(\lambda)$ and to check that the minimal realization has order zero and $\rank D_m = 0$. Numerically reliable methods suitable to compute minimal realizations of descriptor systems perform orthogonal similarity transformations on the system matrices as in (\ref{dssim}), with $Q$ and $Z$ orthogonal transformation matrices, to remove uncontrollable and unobservable finite and infinite eigenvalues and then remove non-dynamic modes using residualization formulas. This approach is described in details in \cite[Section 10.3.1]{Varg17} and  relies on numerically stable algorithms proposed in \cite{Varg90} to compute irreducible (i.e., controllable and observable) realizations of descriptor systems.

The potential weakness of this approach is the need to use thresholds for the rank decisions underlying the reduction algorithms in \cite{Varg90}.
The choice of suitable thresholds, which allow to perform all rank decision correctly, is not
straightforward, and often it is even not possible to use a single threshold such that all rank decisions are correctly performed. Therefore, for large order systems this method can fail to produce the expected order reduction, and therefore must be cautiously used  to assess null rank conditions. It could be advantageous to use this approach in conjunction with norm-based methods (see later) to increase the overall efficiency and reliability.

The minimal realization approach described in \cite[Section 10.3.1]{Varg17} underlies the implementation of the function \textbf{\texttt{gminreal}} of \textbf{DSTOOLS}. If \texttt{G} contains the system object for the rational matrix $G(\lambda)$ with its descriptor system realization of the form (\ref{GTFMalt}) and if \texttt{tol} is a given threshold for rank computations, then the zero normal rank of $G(\lambda)$ can be assessed using the MATLAB commands

\begin{verbatim}

     Gr = gminreal(G,tol);
     isnullrank = (order(Gr) == 0 && rank(Gr.d,tol) == 0)

\end{verbatim}

\emph{Note:} The minimal realization function \textbf{\texttt{minreal}} available in the Control System Toolbox of MATLAB is, in general, not applicable to improper descriptor system realizations. Even for proper systems, this function often fails to reduce the order to zero, independently of the employed tolerance for rank determinations. The problem lies in the employed minimal realization approach, which strongly focuses on preserving the magnitude of the system frequency response. For a system with zero frequency response, this approach automatically leads to an ill-posed problem and therefore, often the expected order reduction does not take place.
%To improve the behaviour of \textbf{\texttt{minreal}} it is possible to add and subtract a, say first order stable transfer function matrix $G_1(\lambda)$ and assess null rank using the following sequence
%\begin{verbatim}
%
%     G1 = drss(1,size(G,1),size(G,2)); G1.a = 0.95*G1.a;
%     Gr = minreal(G+G1,tol);
%     isnullrank = (order(Gr) == 1 && norm(Gr-G1,inf) < tol)
%
%\end{verbatim}

\subsection{Method 2: Norm-based assessment of zero normal rank}
For a proper descriptor system without eigenvalues on the boundary of stability domain $\mathds{C}_s$\footnote{For a continuous-time system, $\mathds{C}_s$ is the open left hand complex plane and its boundary is the extended imaginary axis (i.e., including also the point at infinity), while for a discrete-time system, $\mathds{C}_s$ is the open unit disc centered in the origin and its boundary is the unit circle centered in the origin.}, the zero rank condition  $\rank G(\lambda) = 0$ is equivalent to the $\mathcal{L}_\infty$-norm based condition  $\|G(\lambda)\|_\infty = 0$. This approach is not applicable if the system is not proper in a continuous-time setting, or if there exist eigenvalues on the boundary of $\mathds{C}_s$. In both these cases the norms are by definition infinite. To make this approach generally applicable to arbitrary $G(\lambda)$, it is possible to use  a suitable bilinear transformation $\lambda = g(\delta):= \frac{a\delta+b}{c\delta+d}$, such that the realization $(\widetilde A-\lambda \widetilde E,\widetilde B,\widetilde C,\widetilde D)$ of $G(g(\delta))$ has shifted eigenvalues and the pair $\widetilde A,\widetilde E)$ has only finite eigenvalues.
Clearly, $G(\lambda) = 0$ if $\|G(g(\delta))\|_\infty = 0$. For most cases, the real parameters $a$, $b$, $c$ and $d$ can be simply chosen at random. For proper systems with non-singular $E$, polynomial $g(\delta)$ of the form $g(\delta) = a\delta + b$ can be employed, which simultaneously achieves eigenvalue shifting (if $b\neq 0$) and eigenvalue scaling (if $|a| > 0$).

The descriptor realization $(\widetilde A-\delta \widetilde E,\widetilde B,\widetilde C,\widetilde D)$ for $g(\delta):= \frac{a\delta+b}{c\delta+d}$ can be determined without involving matrix inversions as follows:
\[ \widetilde A-\delta \widetilde E = \ba{cc} dA-bE -\delta (aE-cA) & dB+\delta cB\\
0 & -I \ea, \quad \widetilde B = \ba{c} 0\\ I \ea, \quad \widetilde C = [\, C \; D\,], \quad \widetilde D = 0 . \]
In the polynomial case $g(\delta) = a\delta + b$ with $E$ non-singular, simpler expressions can be used
\[ \widetilde A-\delta \widetilde E = A-bE -\delta aE, \quad \widetilde B = B, \quad \widetilde C = C, \quad \widetilde D = D . \]

The above formulas are implemented in the function \textbf{\texttt{gbilin}} available in \textbf{DSTOOLS}. If \texttt{G} contains the system object for the rational matrix $G(\lambda)$ with its descriptor system realization of the form (\ref{GTFMalt}) and if \texttt{tol} is a given threshold for the zero $\mathcal{L}_\infty$-norm, then the zero normal rank of $G(\lambda)$ can be generally assessed using the MATLAB command

\begin{verbatim}

     isnullrank = (norm(gbilin(G,tf(rand(1,2),rand(1,2),G.Ts)),inf) < tol)

\end{verbatim}

\subsection{Method 3: Direct computation of normal rank}
The most natural approach to check the null rank condition $\rank G(\lambda) = 0$ is to directly evaluate the normal rank of $G(\lambda)$ and check it ii is zero. This can be done using the equivalent relation
\be\label{nrank} \rank G(\lambda) = \rank S(\lambda) - n ,\ee
where
\[ S(\lambda) := \ba{cc} A-\lambda E & B \\ C & D \ea \]
is the so-called system matrix pencil. To evaluate the rank of $S(\lambda)$, it is possible to reduce, using two orthogonal transformation matrices $Q$ and $Z$,  the system matrix pencil $S(\lambda)$ to a Kronecker-like form
\[ \widetilde S(\lambda) = Q^T S(\lambda) Z =
\ba{ccc} A_r-\lambda E_r  & \ast & \ast \\
 0 & A_{reg}-\lambda E_{reg}  & \ast \\
 0 & 0 & A_l-\lambda E_l\ea ,\]
where $A_r-\lambda E_r$ is a $m_r\times n_r$ full row rank pencil,   $A_{reg}-\lambda E_{reg}$ is a $n_{reg}\times n_{reg}$ regular pencil, and $A_l-\lambda E_l$ is a $m_l\times n_l$ full column rank pencil. The generalized eigenvalues of the pair $(A_{reg},E_{reg})$ are the eigenvalues of the system matrix pencil $S(\lambda)$ (also called invariant zeros).
The normal rank of $S(\lambda)$ is thus $m_r+n_{reg}+n_l$  and we have
\[   \rank G(\lambda) = m_r+n_{reg}+n_l -n .\]

Two categories of algorithms can be used to reduce a linear pencil to a Kronecker-like form, which mainly differ in the employed rank estimation technique and the worst-case numerical complexity of the involved computations. The first category of methods uses the singular value decomposition (SVD) as basis for rank determinations, by performing SVD-based row and column compressions. Algorithms in this category have been proposed in \cite{Door79a,Demm93}. Albeit numerically reliable, these algorithms have a worst-case computational complexity $\mathcal{O}(n^4)$, which for large system orders are usually not acceptable.  More efficient algorithms of complexity $\mathcal{O}(n^3)$ have been proposed in \cite{Beel88,Varg96d,Oara97}, which rely on using QR decompositions with column pivoting for rank determinations, and for row and column compressions. An alternative approach tailored to the structure of a system matrix pencil has been proposed in \cite{Emam82} for standard systems (i.e., $E = I_n$), and in \cite{Misr94} for general descriptor systems. The exclusive use of orthogonal transformations to perform various reductions make these methods numerically stable, ensuring that the computed reduced pencil corresponds to slightly perturbed system data in the range of thresholds used for rank decisions.

Unfortunately, the need to use of thresholds for rank decisions represents the main weakness of these algorithms.   The choice of suitable thresholds, which allow to perform all rank decision correctly, is not straightforward, and often it is not possible to use a single threshold such that all rank decisions are correct. Therefore, for large order systems all these methods must be cautiously used to assess null rank conditions.

The algorithms of \cite{Emam82} and \cite{Misr94} form the basis of the computational method underlying the function \textbf{\texttt{gnrank}} of \textbf{DSTOOLS}. If \texttt{G} contains the system object for the rational matrix $G(\lambda)$ or for its descriptor system realization and if \texttt{tol} is a given threshold for rank computations, then the normal rank of $G(\lambda)$ can be assessed using the MATLAB command

\begin{verbatim}

     isnullrank = (gnrank(G,tol) == 0);

\end{verbatim}

\subsection{Method 4: Direct estimation of normal rank from the frequency response}
Since the normal rank of $G(\lambda)$ is the maximally possible rank of the complex matrix $G(\lambda)$ for all values of $\lambda \in \mathds{C}$ such that $G(\lambda)$ has finite norm, it is possible to estimate the normal rank as the maximum of the rank of $G(\lambda)$ for a few random values of the frequency variable $\lambda$. Thus, if $\Omega$ contains a set of complex frequency values, which are neither poles nor zeros of $G(\lambda)$, then the normal rank $r$ can be estimated as
\[ r = \max_{\lambda_z \in \Omega} \rank G(\lambda_z) , \]
where $G(\lambda_z)$ is computed as
\[ G(\lambda_z) = C(\lambda_z E-A)^{-1}B+D . \]
In practice, a single random value of $\lambda_z$ is usually sufficient to estimate the correct rank of $G(\lambda)$. For a more reliable rank estimation, a few more values can be used.

The main appeal of this method  is its high computational efficiency even for large values of $n$. Also, the method is quite reliable in estimating the correct rank, provided the chosen frequency values are sufficiently far away from the finite generalized eigenvalues of the pair $(A,E)$ and from the eigenvalues of the system matrix pencil $S(\lambda)$.

If \texttt{G} contains the system object for the rational matrix $G(\lambda)$ or for its descriptor system realization and if \texttt{tol} is a given threshold for rank computations, then the null normal rank of $G(\lambda)$ can be assessed with this method using the MATLAB command

\begin{verbatim}

     isnullrank = (rank(evalfr(G,rand),tol) == 0);

\end{verbatim}

\subsection{Method 5: Estimation of the normal rank from the normal rank of the system matrix pencil}
We can alternatively use (\ref{nrank}) to estimate the normal rank as
\be\label{nrest} r = \max_{\lambda_z \in \Omega} \rank S(\lambda_z) - n , \ee
where $\Omega$ contains a set of complex frequency values which are neither poles and zeros of $G(\lambda)$. This approach avoids all numerical difficulties related to choosing frequency values, as nearby values to the eigenvalues of $A-\lambda E$ or of to the eigenvalues of the system matrix pencil $S(\lambda)$. Also, choosing a suitable threshold for rank decisions is here alleviated by the fact that $S(\lambda_z)$ has usually nonzero rank, thus the use default tolerances based on norm estimations is reliable.

The estimation of the normal rank based on (\ref{nrest}) is implemented as an option of the function \textbf{\texttt{gnrank}} of \textbf{DSTOOLS}.
If \texttt{G} contains the system object for the descriptor realization of the rational matrix $G(\lambda)$, then the normal rank of $G(\lambda)$ can be estimated using the MATLAB commands

\begin{verbatim}

     isnullrank = (gnrank(G,tol,rand) == 0);

\end{verbatim}

\section{Numerical example}\label{sec:examples} To illustrates the capabilities of the discussed approaches, we consider a simple example involving descriptor systems models of increasing orders. These models originate from a standard operation on standard systems performed with functions of the Control System Toolbox of MATLAB and of the \textbf{DSTOOLS} collection. Consider an $n$-th order discrete-time system with 2 outputs and 3 inputs with the proper \textbf{transfer} function matrix $R(z)$. We form $G(z)$ as the difference $G(z) = R^\sim(z) -R^\sim(z)$, with the conjugated system $R^\sim(z) =: R^T(1/z)$ computed in two ways: directly, using the pertransposition function of the Control System Toolbox of MATLAB and, in two steps, using the bilinear transformation $z \rightarrow 1/z$ and the transpose function of the Control System Toolbox of MATLAB.  The realization of $G(z)$ is obtained for a given value of $n$ using the MATLAB commands

\begin{verbatim}

     R = drss(n,2,3);
     G = R' - gbilin(R,tf(1,[1 0],-1)).';

\end{verbatim}

The resulting descriptor system realization of $G(z)$ has order $N = 2n+2$, and has 2 non-dynamic modes, $n$ finite uncontrollable eigenvalues and $n$ finite unobservable eigenvalues.

For values of $n = \{ 1, 2, 3, 5, 10, 20, 50, 100, 200 \}$ we performed the assessment of the null rank condition with the previously described five methods. We used a tolerance \texttt{tol = 1.e-7} for both rank determinations and norm-based evaluations.  In Table \ref{t1} we present for each value of $n$, the resulting null rank decisions (parameter \texttt{isnullrank} is set 1 for \texttt{true} or 0 for \texttt{false}) for each method. The \textbf{Method 1} and \textbf{Method 3} use the single threshold value \texttt{tol} for all rank decisions intervening in  computing many row and column compressions. As it can be seen from Table \ref{t1}, these methods, although relying on numerically stable computations, produce wrong rank evaluations for order exceeding 20 and therefore are not recommended to be used for large order systems. The norm computation based \textbf{Method 2} and the rank estimation based methods \textbf{Method 4} and \textbf{Method 5} perform equally well on this examples, even for relatively large system orders. In particular, the \textbf{Method 5} is able to correctly assess the normal rank even using the default setting of the threshold used for rank determinations.
\begin{table}[!h]
\begin{center}
\begin{tabular}{ccccccc}
$n$ & $N$ & \textbf{Method 1} & \textbf{Method 2} & \textbf{Method 3} & \textbf{Method 4} & \textbf{Method 5} \\ \hline
     1 &    4  &   1  &   1  &   1  &   1  &   1\\
     2 &    6  &   1  &   1  &   1  &   1  &   1\\
     3 &    8  &   1  &   1  &   1  &   1  &   1\\
     5 &   12  &   1  &   1  &   1  &   1  &   1\\
    10 &   22  &   1  &   1  &   1  &   1  &   1\\
    20 &   42  &   1  &   1  &   1  &   1  &   1\\
    50 &  102  &   0  &   1  &   0  &   1  &   1\\
   100 &  202  &   0  &   1  &   0  &   1  &   1\\
   200 &  402  &   0  &   1  &   0  &   1  &   1
\end{tabular}
\end{center}
\caption{Results obtained for checking the null rank conditions}  \label{t1}
\end{table}

In Table \ref{t2}, we give the total elapsed times necessary for each method to address the rank determinations for all values of $n$.  Of the three methods of choice (i.e., \textbf{Method 2}, \textbf{Method 4} and \textbf{Method 5}), the fastest one is \textbf{Method 4}, but \textbf{Method 5} has a comparable speed, while \textbf{Method 2} is substantially slower.
\begin{table}[!h]
\begin{center}
\begin{tabular}{ccccc}
\textbf{Method 1} & \textbf{Method 2} & \textbf{Method 3} & \textbf{Method 4} & \textbf{Method 5} \\ \hline
 0.75 &   4.31 &   0.25 &   0.01 &   0.04
\end{tabular}
\end{center}
\caption{Total elapsed times (in seconds) for checking the null rank conditions}\label{t2}
\end{table}

On the basis of our experimental evaluations, also performed on similar problems, \textbf{Method 5} appears to be the best suited for the assessment of null rank conditions of arbitrary rational matrices.

\section{Conclusions}
\label{sec:conclusions}

In this paper we surveyed five numerical approaches to check null rank conditions of rational transfer function matrices of continuous- or discrete-time descriptor systems. The best numerical performance (i.e., numerical reliability, computational efficiency) exhibits \textbf{Method  5}, which  is based on the estimation of the normal rank of the system matrix pencil. The only computation is the evaluation of the rank of this pencil for a suitably chosen value of the frequency variable  (or, for increased reliability, for a few number of values of the frequency variable). This rank determination is reliably performed using the singular value decomposition, which also allows to employ easily computable default values of the rank decision threshold. The Method 5 to estimate the normal rank of a transfer function matrix is implemented in the function \textbf{\texttt{gnrank}} of the \textbf{DSTOOLS} collection \cite{Varg18}.

%\appendix
%\section{An example appendix}
%\lipsum[71]
%
%\section*{Acknowledgments}
%We would like to acknowledge the assistance of volunteers in putting
%together this example manuscript and supplement.

\bibliographystyle{siamplain}
%\bibliography{../../varga}

\end{document}